\documentclass{article}
\usepackage{amsfonts}
\usepackage{amsmath}

\input{tcilatex}
\begin{document}

\title{Path integral solution for a deformed radial Rosen-Morse potential}
\author{A Kadja, F Benamira and L Guechi* \\
%EndAName
Laboratoire de Physique Th\'{e}orique, D\'{e}partement de Physique, \and %
Facult\'{e} des Sciences Exactes, Universit\'{e} des fr\`{e}res Mentouri,
\and Route d'Ain El Bey, Constantine, Algeria.}
\maketitle

\begin{abstract}
An exact path integral treatment of a particle in a deformed radial \
Rosen-Morse potential is presented. For this problem with the Dirichlet
boundary conditions, the Green's \ function is constructed in a closed form
by adding to $V_{q}(r)$ a $\delta -$function perturbation and making its
strength infinitely repulsive. A transcendental equation for the energy
levels $E_{n_{r}}$ and the wave functions of the bound states can then be
deduced.


03.65.Ca-Formalism

03.65.Db-Functional analytical methods


*Corresponding author, E-mail: guechilarbi@yahoo.fr
\end{abstract}

\section{Introduction}

The so-called Rosen-Morse potential is a one-dimensional potential function
introduced by Rosen and Morse in 1932 to study the vibrational states of
poly-atomic molecules \cite{Rosen} . Since then, it has attracted a lot of
interest due to its numerous applications in several branches of physics 
\cite{Townes,Smondyrev}. It has also been used as an illustrative example in
different methods such as the factorization method \cite{Infeld,Amani}, the
prepotential approach \cite{Ho}, the path integral technique \cite%
{Junker,Grosche,Mustapic,Kleinert}, the supersymmetry in quantum mechanics
and the shape invariance \cite{Cooper,Yilmaz} and the Nikiforov-Uvarov
method \cite{Egrifes1,Egrifes2,Egrifes3}.

There is also the spherically symmetric Rosen-Morse potential which has been
discussed by many authors \cite{Taskin,Jia,Diaf1,Diaf2,Antia} in recent
years without distinguishing between the one-dimensional potential and the
radial potential problem. Consequently, the solutions which have been
obtained until now are not satisfactory.

The purpose of the present work is to re-examine and rederive, within the
framework of path integrals, the correct solutions of the problem of the non
relativistic particle of mass $M$ moving in the $q-$deformed radial
Rosen-Morse potential denoted by 
\begin{equation}
V_{q}\left( r\right) =-\frac{V_{1}}{\cosh _{q}^{2}\left( \frac{r}{a}\right) }%
+V_{2}\tanh _{q}\left( \frac{r}{a}\right)  \label{RM.1}
\end{equation}%
where $V_{1},V_{2},a$ and $q$ are four potential parameters. It is defined
in terms of the $q-$deformed hyperbolic functions $(q>0)$

\begin{equation}
\sinh _{q}x=\frac{e^{x}-qe^{-x}}{2},\text{ \ }\cosh _{q}x=\frac{e^{x}+qe^{-x}%
}{2},\text{ \ }\tanh _{q}x=\frac{\sinh _{q}x}{\text{\ }\cosh _{q}x}
\label{RM.1a}
\end{equation}%
which have been introduced for the first time by Arai \cite{Arai}. The
introduction of the parameter $q$ may be used as an additional parameter in
the description of inter-atomic interactions, in particular, when the
particle motion takes place in a half-space different from the half-space $%
r>0$, i. e., the center of mass location of the molecule is not at the
coordinate origin. In Fig. \ref{Fig.1} the potential (\ref{RM.1}) is plotted
for four different $q$ values.

\begin{equation*}
\FRAME{itbpFU}{2.7095in}{2.0686in}{0in}{\Qcb{Fig. \protect\ref{Fig.1}: A
plot of the Rosen-Morse potential (\protect\ref{RM.1}) for different $q$
values. Here $V_{2}=V_{1}/4$. The potential $V_{q}\left( r\right) $ is
scaled in units of $V_{1}$, and the coordinate $r$ is scaled in units of $a.$%
}}{\Qlb{Fig.1}}{swp11.jpg}{\special{language "Scientific Word";type
"GRAPHIC";maintain-aspect-ratio TRUE;display "USEDEF";valid_file "F";width
2.7095in;height 2.0686in;depth 0in;original-width 2.6671in;original-height
2.0306in;cropleft "0";croptop "1";cropright "1";cropbottom "0";filename
'swp11/swp11.jpg';file-properties "NPEU";}}
\end{equation*}

To do that, in section 2, the evaluation procedure of the Green's function
via the path integral approach for a potential with Dirichlet boundary
conditions is used. We use a trick which consists in incorporating a $\delta
-$ function perturbation as an additional potentiel. \ After calculating the
Green's function in closed form, i.e., in a form that involves no summation,
we shall make the strength of the $\delta -$ function perturbation
infinitely repulsive to obtain the Green's function associated with the $s$
waves for the potential (\ref{RM.1}). Then in section 3, we derive a
transcendental equation for the energy levels and the wave functions of the
bound states. In section 4, the standard radial Rosen-Morse ($q=1$) is
considered as particular case. The section 5 will be a conclusion.

\section{ Green's function}

The propagator for a particle of mass $M$ in the deformed radial Rosen-Morse
(\ref{RM.1}) is written in spherical coordinates as

\begin{equation}
K^{q}\left( \overrightarrow{r}^{\prime \prime },\overrightarrow{r}^{\prime
};T\right) =\frac{1}{r^{\prime \prime }r^{\prime }}\sum\limits_{l=0}^{\infty
}\frac{2l+1}{4\pi }K_{l}^{q}\left( r^{\prime \prime },r^{\prime };T\right)
P_{l}\left( \frac{\overrightarrow{r}^{\prime \prime }.\overrightarrow{r}%
^{\prime }}{r^{\prime \prime }r^{\prime }}\right) ,  \label{RM.2}
\end{equation}%
where $P_{l}\left( \frac{\overrightarrow{r}^{\prime \prime }.\overrightarrow{%
r}^{\prime }}{r^{\prime \prime }r^{\prime }}\right) $ is the Legendre
polynomial of degree $l$ in $\frac{\overrightarrow{r}^{\prime \prime }.%
\overrightarrow{r}^{\prime }}{r^{\prime \prime }r^{\prime }}=\cos \theta
^{\prime \prime }\cos \theta ^{\prime }+\sin \theta ^{\prime \prime }\sin
\theta ^{\prime }\cos (\phi ^{\prime \prime }-\phi ^{\prime })$. The radial
propagator is given by the path integral representation \cite{Peak}

\begin{equation}
K_{l}^{q}\left( r^{\prime \prime },r^{\prime };T\right) =\text{ }\underset{%
N\rightarrow \infty }{\lim }\prod\limits_{j=1}^{N}\left[ \frac{M}{2i\pi
\hbar \varepsilon }\right] ^{\frac{1}{2}}\prod\limits_{j=1}^{N-1}\left[ \int
dr_{j}\right] \exp \left[ \frac{i}{\hbar }\sum\limits_{j=1}^{N}S(j,j-1)%
\right] ,  \label{RM.3}
\end{equation}%
in which the action for each short-time interval is defined by

\begin{equation}
S(j,j-1)=\frac{M}{2\varepsilon }\left( \Delta r_{j}\right) ^{2}+\varepsilon 
\frac{l(l+1)\hbar ^{2}}{2Mr_{j}r_{j-1}}+\varepsilon \left( \frac{V_{1}}{%
\cosh _{q}^{2}\left( \frac{r_{j}}{a}\right) }-V_{2}\tanh _{q}\left( \frac{%
r_{j}}{a}\right) \right) ,  \label{RM.4}
\end{equation}%
with the usual notation $r_{j}=r(t_{j}),\triangle
r_{j}=r_{j}-r_{j-1},\varepsilon =t_{j}-t_{j-1}=T/N,t^{\prime
}=t_{0},t^{\prime \prime }=t_{N}$ and $T=t^{\prime \prime }-t^{\prime }$. By
assuming that the system has only a discrete spectrum, expression (\ref{RM.4}%
) corresponds to the propagator expressed in spectral expansion as

\begin{equation}
K_{l}^{q}\left( r^{\prime \prime },r^{\prime };T\right)
=\sum\limits_{n_{r}}\varphi _{n_{r},l}^{q\ast }(r^{\prime })\varphi
_{n_{r},l}^{q}(r^{\prime \prime })e^{\frac{i}{\hbar }E_{n_{r},l}T};\text{ }%
T>0,  \label{RM.5}
\end{equation}%
where $\varphi _{n_{r},l}^{q}(r)$ is the reduced radial wave function, $%
E_{n_{r},l}$ are the energy eigenvalues and $n_{r}$ denotes the number of
the nodes of the radial wave functions. Our aim is to find $E_{n_{r}}$ and $%
\varphi _{n_{r}}^{q}(r)$ for $l=0$ by evaluating (\ref{RM.3}). Since the
radial path integral (\ref{RM.3}) cannot directly be calculated, we consider
the radial Green's function (Fourier transform of the radial propagator):

\begin{equation}
G_{0}^{q}\left( r^{\prime \prime },r^{\prime };E\right) =\int_{0}^{\infty
}dT\exp \left( \frac{i}{\hbar }ET\right) K_{0}^{q}\left( r^{\prime \prime
},r^{\prime };T\right) .  \label{RM.6}
\end{equation}%
With the new variable $u=\frac{r}{a}-\ln \sqrt{q}$ and the new pseudo-time $%
s=\frac{t}{a^{2}}$, \ the radial Green's function (\ref{RM.6}) becomes%
\begin{equation}
G_{0}^{q}\left( r^{\prime \prime },r^{\prime };E\right) =a\widetilde{G}%
_{RM}\left( u^{\prime \prime },u^{\prime };E\right)  \label{RM.7}
\end{equation}%
and%
\begin{equation}
\widetilde{G}_{RM}\left( u^{\prime \prime },u^{\prime };E\right)
=\int_{0}^{\infty }dS\exp \left( \frac{i}{\hbar }a^{2}ES\right)
K_{0}(u^{\prime \prime },u^{\prime };S),  \label{RM.7bis}
\end{equation}%
where%
\begin{equation}
K_{0}(u^{\prime \prime },u^{\prime };S)=\int Du(s)\exp \left\{ \frac{i}{%
\hbar }\int_{0}^{S}\left( \frac{M}{2}\overset{.}{u}^{2}-V(u)\right)
ds\right\} ,  \label{RM.8}
\end{equation}%
with 
\begin{equation}
V(u)=a^{2}\left( V_{2}\tanh u-\frac{V_{1}}{q\cosh ^{2}u}\right) ;\text{ \ \ }%
u\geq -\ln \sqrt{q}.  \label{RM.9}
\end{equation}%
The propagator (\ref{RM.8}) has the same form as the path integral
associated with the Rosen-Morse potential $V_{RM}(u)$ for $u\in 
%TCIMACRO{\U{211d} }%
%BeginExpansion
\mathbb{R}
%EndExpansion
$, but, in the present case, we have converted the path integral for the
deformed radial potential (\ref{RM.1}) into a path integral for a
Rosen-Morse potential type by means of the transformation $r\rightarrow r(u)$%
, which maps $%
%TCIMACRO{\U{211d} }%
%BeginExpansion
\mathbb{R}
%EndExpansion
^{+}\rightarrow \left] -\ln \sqrt{q},+\infty \right[ $. This means that the
motion of the particle takes place on the half line $u\geq u_{0}=-\ln \sqrt{q%
}.$ As a direct path integration is not possible, the problem can be solved
with the help of a trick which consists in adding an auxiliary $\delta -$%
function term to the action contained in Eq. (\ref{RM.7bis}) to form an
impenetrable wall \cite{Clark} at $u=u_{0}=-\ln \sqrt{q}$ by letting the
strength of the $\delta -$function be infinitely repulsive. Then, in this
case, the Green's function is given by 
\begin{equation}
G_{0}^{\delta }\left( u^{\prime \prime },u^{\prime };E\right)
=\int_{0}^{\infty }dS\exp \left( \frac{i}{\hbar }a^{2}ES\right)
K_{0}^{\delta }(u^{\prime \prime },u^{\prime };S)  \label{RM.10}
\end{equation}%
where%
\begin{equation}
K_{0}^{\delta }(u^{\prime \prime },u^{\prime };S)=\int Du(s)\exp \left\{ 
\frac{i}{\hbar }\int_{0}^{S}\left( \frac{M}{2}\overset{.}{u}^{2}-V^{\delta
}(u)\right) ds\right\}  \label{RM.11}
\end{equation}%
is the propagator for a one-dimensional system bounded by a potential of the
form: 
\begin{equation}
V^{\delta }(u)=V_{RM}(u)-\lambda \delta \left( u-u_{0}\right) ;\text{ \ \ \ }%
u\in 
%TCIMACRO{\U{211d} }%
%BeginExpansion
\mathbb{R}
%EndExpansion
,  \label{RM.12}
\end{equation}%
in which $V_{RM}(u)$ is the standard Rosen-Morse potential. As the path
integration of (\ref{RM.11}) cannot directly be performed, we apply the
perturbative approach by expanding $\exp \left( \frac{i}{\hbar }\lambda
\int_{s^{\prime }}^{s^{\prime \prime }}\delta \left( u-u_{0}\right)
ds\right) $ into the power series. This gives the following series expansion 
\cite{Grosche1,Hibbs,Lawande} :

\begin{eqnarray}
K_{0}^{\delta }(u^{\prime \prime },u^{\prime };S) &=&K_{RM}(u^{\prime \prime
},u^{\prime };S)  \notag \\
&&+\underset{n=1}{\overset{\infty }{\sum }}\frac{1}{n!}\left( \frac{i}{\hbar 
}\lambda \right) ^{n}\int \mathit{D}u(s)\exp \left[ \frac{i}{\hbar }%
\int_{0}^{S}\left( \frac{M}{2}\overset{.}{u}^{2}-V_{RM}(u)\right) ds\right] 
\notag \\
&&\times \int_{0}^{S}\delta \left( u_{1}-u_{0}\right)
ds_{1}...\int_{0}^{S}\delta \left( u_{n}-u_{0}\right) ds_{n}  \label{RM.13}
\end{eqnarray}%
where $K_{RM}(u^{\prime \prime },u^{\prime };S)$ is the propagator for the
standard Rosen-Morse potential. With the time-ordering $s^{\prime
}=s_{0}=0<s_{1}<s_{2}<...<s_{n}<s^{\prime \prime }=S$, the propagator (\ref%
{RM.13}) can be rewritten as the Feynman-Dyson perturbation series 
\begin{eqnarray}
K_{0}^{\delta }(u^{\prime \prime },u^{\prime };S) &=&K_{RM}(u^{\prime \prime
},u^{\prime };S)  \notag \\
&&+\underset{n=1}{\overset{\infty }{\sum }}\frac{1}{n!}\left( \frac{i}{\hbar 
}\lambda \right) ^{n}\overset{n}{\underset{j=1}{\prod }}\left[
\int_{s^{\prime }}^{s_{j+1}}ds_{j}\int_{-\infty }^{+\infty }du_{j}\right] 
\notag \\
&&\times K_{RM}(u_{1},u^{\prime };s_{1}-s^{\prime })\delta \left(
u_{1}-u_{0}\right) K_{RM}(u_{2},u_{1};s_{2}-s_{1})  \notag \\
&&\times ...\times \delta \left( u_{n-1}-u_{n-2}\right)
K_{RM}(u_{n},u_{n-1};s_{n}-s_{n-1})  \notag \\
&&\times \delta \left( u_{n}-u_{n-1}\right) K_{RM}(u^{\prime \prime
},u_{n};s^{\prime \prime }-s_{n})  \notag \\
&=&K_{RM}(u^{\prime \prime },u^{\prime };S)+\underset{n=1}{\overset{\infty }{%
\sum }}\frac{1}{n!}\left( \frac{i}{\hbar }\lambda \right)
^{n}\int_{s^{\prime }}^{s^{\prime \prime }}ds_{n}  \notag \\
&&\times \int_{s^{\prime }}^{s_{n}}ds_{n-1}...\int_{s^{\prime
}}^{s_{2}}ds_{1}K_{RM}(u_{1},u^{\prime };s_{1}-s^{\prime })  \notag \\
&&\times K_{RM}(u_{2},u_{1};s_{2}-s_{1})\times ...\times K_{RM}(u^{\prime
\prime },u_{n};s^{\prime \prime }-s_{n}).  \notag \\
&&  \label{RM.14}
\end{eqnarray}%
In order to perform the successive integrations over the time variables $%
s_{j}$ in (\ref{RM.14}), we insert (\ref{RM.14}) into (\ref{RM.10}), and
making use the convolution theorem of the Fourier transformation, we arrive
at 
\begin{equation}
G_{RM}^{\delta }(u^{\prime \prime },u^{\prime };E)=G_{RM}(u^{\prime \prime
},u^{\prime };E)-\frac{G_{RM}(u^{\prime \prime
},u_{0};E)G_{RM}(u_{0},u^{\prime };E)}{G_{RM}(u_{0},u_{0};E)-\frac{1}{%
\lambda }},  \label{RM.15}
\end{equation}%
where $G_{RM}(u^{\prime \prime },u^{\prime };E)$ is the Green's function
relative to the Rosen-Morse potential in the entire space $%
%TCIMACRO{\U{211d} }%
%BeginExpansion
\mathbb{R}
%EndExpansion
$, and, as is known from the literature \cite{Mustapic,Kleinert}, its closed
expression is given by

\begin{eqnarray}
G_{RM}(u^{\prime \prime },u^{\prime };E) &=&-\frac{iM}{\hbar }\frac{\Gamma
(M_{1}-L_{\nu _{q}})\Gamma (L_{\nu _{q}}+M_{1}+1)}{\Gamma
(M_{1}+M_{2}+1)\Gamma (M_{1}-M_{2}+1)}  \notag \\
&&\times \left( \frac{1-\tanh u^{\prime }}{2}.\frac{1-\tanh u^{\prime \prime
}}{2}\right) ^{\frac{M_{1}+M_{2}}{2}}  \notag \\
&&\times \text{ }\left( \frac{1+\tanh u^{\prime }}{2}.\frac{1+\tanh
u^{\prime \prime }}{2}\right) ^{\frac{M_{1}-M_{2}}{2}}  \notag \\
&&\times \text{ }_{2}F_{1}\left( M_{1}-L_{\nu _{q}},L_{\nu
_{q}}+M_{1}+1,M_{1}-M_{2}+1;\frac{1}{2}(1+\tanh u_{>})\right)  \notag \\
&&\times \text{ }_{2}F_{1}\left( M_{1}-L_{\nu _{q}},L_{\nu
_{q}}+M_{1}+1,M_{1}+M_{2}+1;\frac{1}{2}(1-\tanh u_{<})\right) ,  \notag \\
&&  \label{RM.16}
\end{eqnarray}%
with the following abbreviations%
\begin{equation}
\left\{ 
\begin{array}{c}
L_{\nu _{q}}=\frac{1}{2}\left( \nu _{q}-1\right) , \\ 
\nu _{q}=\sqrt{1+\frac{8Ma^{2}V_{1}}{\hbar ^{2}q}}, \\ 
M_{1,2}=\sqrt{\frac{Ma^{2}}{2\hbar ^{2}}}\left( \sqrt{V_{2}-E}\pm \sqrt{%
-V_{2}-E}\right) .%
\end{array}%
\right.  \label{RM.17}
\end{equation}%
The $_{2}F_{1}(\alpha ,\beta ,\gamma ,u)$ are the hypergeometric functions.
The symbols $u_{>}$ and $u_{<}$ denote $\max (u^{\prime \prime },u^{\prime
}) $ and $\min (u^{\prime \prime },u^{\prime })$, respectively.

Now, in the limit $\lambda \rightarrow -\infty $, the physical system is
forced to move in the potential $V_{RM}(u)$ bounded by an infinitely
repulsive barrier \cite{Clark,Grosche1} located at $u=u_{0}$. In this case,
the Green's function is given by

\begin{eqnarray}
\widetilde{G}_{RM}(u^{\prime \prime },u^{\prime };E) &=&\underset{\lambda
\rightarrow -\infty }{\lim }G_{RM}^{\delta }(u^{\prime \prime },u^{\prime
};E)  \notag \\
&=&G_{RM}(u^{\prime \prime },u^{\prime };E)-\frac{G_{RM}(u^{\prime \prime
},u_{0};E)G_{RM}(u_{0},u^{\prime };E)}{G_{RM}(u_{0},u_{0};E)}  \notag \\
&&  \label{RM.18}
\end{eqnarray}

\section{Energy spectrum and wave functions of bound states}

The energy spectrum for the bound states can be obtained from the poles of
the Green's function (\ref{RM.18}), i.e. by the equation $%
G_{RM}(u_{0},u_{0};E)=0$, or as well by the transcendental equation%
\begin{equation}
\text{ }_{2}F_{1}\left( M_{1}(E_{n_{r}})-L_{\nu _{q}},L_{\nu
_{q}}+M_{1}(E_{n_{r}})+1,M_{1}(E_{n_{r}})+M_{2}(E_{n_{r}})+1;\frac{q}{q+1}%
\right) =0.  \label{RM.19}
\end{equation}%
The energies $E_{n_{r}}$ of the bound states can be determined by solving
numerically this equation and the wave functions satisfying the boundary
conditions $\varphi _{n_{r}}^{q}(0)=\varphi _{n_{r}}^{q}(+\infty )=0$ are
given by.

\begin{eqnarray}
\varphi _{n_{r}}^{q}(r) &=&C\text{ }\left( \frac{q}{e^{r/a}+q}\right) ^{%
\frac{M_{1}(E_{n_{r}})+M_{2}(E_{n_{r}})}{2}}\left( \frac{e^{2r/a}}{e^{r/a}+q}%
\right) ^{\frac{M_{1}(E_{n_{r}})-M_{2}(E_{n_{r}})}{2}}  \notag \\
&&\times \text{ }_{2}F_{1}\left( M_{1}(E_{n_{r}})-L_{\nu _{q}},L_{\nu
_{q}}+M_{1}(E_{n_{r}})+1,M_{1}(E_{n_{r}})+M_{2}(E_{n_{r}})+1;\frac{q}{%
e^{2r/a}+q}\right) ,  \notag \\
&&  \label{RM.20}
\end{eqnarray}%
where $C$ is a constant factor.

\section{Radial Rosen-Morse potential}

By setting $q=1$ in the expression (\ref{RM.1}), we obtain the so-called
radial Rosen-Morse potential

\begin{equation}
V(r)=-\frac{V_{1}}{\cosh ^{2}\left( \frac{r}{a}\right) }+V_{2}\tanh \left( 
\frac{r}{a}\right) .  \label{RM.21}
\end{equation}

The $L_{\nu _{q}}$ and $\nu _{q}$ parameters defined by expressions (\ref%
{RM.17}) can be written 
\begin{equation}
\left\{ 
\begin{array}{c}
L_{\nu _{1}}=\frac{1}{2}\left( \nu _{1}-1\right) , \\ 
\nu _{1}=\sqrt{1+\frac{8Ma^{2}V_{1}}{\hbar ^{2}}}.%
\end{array}%
\right.  \label{RM.22}
\end{equation}

In this case, it follows from (\ref{RM.19}) and (\ref{RM.20}) that the
transcendental quantization condition for the bound state energy levels $%
E_{n_{r}}$ and the wave functions are

\begin{equation}
\text{ }_{2}F_{1}\left( M_{1}(E_{n_{r}})-L_{\nu _{1}},L_{\nu
_{1}}+M_{1}(E_{n_{r}})+1,M_{1}(E_{n_{r}})+M_{2}(E_{n_{r}})+1;\frac{1}{2}%
\right) =0,  \label{RM.23}
\end{equation}%
and

\begin{eqnarray}
\varphi _{n_{r}}^{1}(r) &=&C\text{ }\left( \frac{1}{e^{r/a}+1}\right) ^{%
\frac{M_{1}(E_{n_{r}})+M_{2}(E_{n_{r}})}{2}}\left( \frac{e^{2r/a}}{e^{r/a}+1}%
\right) ^{\frac{M_{1}(E_{n_{r}})-M_{2}(E_{n_{r}})}{2}}  \notag \\
&&\times \text{ }_{2}F_{1}\left( M_{1}(E_{n_{r}})-L_{\nu _{1}},L_{\nu
_{1}}+M_{1}(E_{n_{r}})+1,M_{1}(E_{n_{r}})+M_{2}(E_{n_{r}})+1;\frac{1}{%
e^{2r/a}+1}\right) .  \notag \\
&&  \label{RM.24}
\end{eqnarray}

\section{Conclusion}

In this paper we have discussed the path integral problem for a deformed
radial Rosen-Morse potential which is a potential with the Dirichlet
boundary conditions. Our approach shows that the path integral in the
present case is more rigorous in comparison to other methods. The derivation
of the exact Green's function for this potential problem contained in this
study is obtained for the first time. The poles and the residues gave a
transcendental quantization condition involving the hypergeometric function
and the wave functions corresponding to the energy levels $E_{n_{r}}$ of the 
$s-$states. The transcendental equation can be solved numerically.

\end{document}